\documentclass[a4paper,twoside]{article}
\usepackage{deluxetable}
\newcommand{\affil}[1]{$^{\rm #1}$}
\usepackage[authoryear]{natbib}
\bibpunct{ (}{)}{;}{a}{}{,}
\usepackage{graphicx}
\usepackage{epsfig}
\usepackage{epstopdf}
\usepackage{gensymb}
\date{Submitted to Rock Art Research 18 August 2011; this version 28 Sep 2012} 
\newcommand{\kms}{\mbox{km\,s$^{-1}$}}
\def\kms {\ifmmode{{\rm ~km~s}^{-1}}\else{~km~s$^{-1}$}\fi}
\def\lsun {\ifmmode{{\rm ~L}_\odot}\else{~L$_\odot$}\fi}
\def\deg {^{\circ} }

\newbox\grsign \setbox\grsign=\hbox{$>$} \newdimen\grdimen \grdimen=\ht\grsign
\newbox\simlessbox \newbox\simgreatbox
\setbox\simgreatbox=\hbox{\raise.5ex\hbox{$>$}\llap
 {\lower.5ex\hbox{$\sim$}}}\ht1=\grdimen\dp1=0pt
\setbox\simlessbox=\hbox{\raise.5ex\hbox{$<$}\llap
 {\lower.5ex\hbox{$\sim$}}}\ht2=\grdimen\dp2=0pt

\def \etal {\rm ~{\it \etal},~}

\title{\large\bf\flushleft {Wurdi Youang: an Australian Aboriginal stone arrangement with possible solar indications.}}
\author{\parbox{\textwidth}{\flushleft
\vspace{-0.5cm}
{\it
Ray P.\ Norris\affil{1,2},
Cilla Norris\affil{3},
Duane W. Hamacher\affil{1,5},
Reg Abrahams\affil{4}
}
 \\
\vspace{0.4cm}
{\small \affil{1}\,Department of Indigenous Studies, Macquarie University, NSW, 2109, Australia}\\ 
{\small \affil{2}\,CSIRO Astronomy \& Space Science, PO Box 76, Epping, NSW, 1710, Australia\\ 
email: {\tt RayPNorris@gmail.com}}\\
{\small \affil{3}\,Emu Dreaming, PO Box 4335, North Rocks, NSW, 2154, Australia}\\
{\small \affil{4}\,Wathaurong Aboriginal Cooperative, Lot 62, Morgan Street, North Geelong, Vic 3215, Australia}\\
{\small \affil{5}\, Nura Gili Centre for Indigenous Programs, University of New South Wales, Sydney, NSW 2052, Australia}\\ 
}}
\begin{document}
\twocolumn[
\begin{changemargin}{.8cm}{.5cm}
\begin{minipage}{.9\textwidth}
\vspace{-1cm}
\maketitle
%
%

{\bf Abstract: 
Wurdi Youang is an egg-shaped Aboriginal stone arrangement in Victoria, Australia. Here we present a new survey of the site, and show that its major axis is aligned within a few degrees of east-west. We confirm a previous hypothesis that it contains alignments to the position on the horizon of the setting  sun at the equinox and the solstices, and show that two independent sets of indicators are aligned in these directions. We show that these alignments are unlikely to have arisen by chance, and instead the builders of this stone arrangement appear to have deliberately aligned the site on astronomically significant positions. 
}

\medskip{\bf Keywords:} Cultural astronomy --- Aboriginal astronomy --- Aboriginal culture --- stone arragements 

\medskip
\medskip
\end{minipage}
\end{changemargin}

]

\section{Introduction}
\label{intro}

\subsection{Aboriginal Astronomy}
It is well established  that the night sky plays an important role in many Australian Aboriginal cultures
\citep{stanbridge,mountford, haynes, johnson, harney, norrisbook, norris09, norris11}. As well as being associated with traditional songs and ceremonies, the sky is used to regulate calendars,  and mark the time of year when a particular  food source appears. 
The sky also had practical applications for navigation and time keeping  \citep{harney, clarke97}, and there is evidence of a search for meaning in astronomical phenomena such as eclipses, planetary motions and tides \citep{norris09}. Astronomical themes are also widespread in ceremonies and artefacts, such as the Morning Star pole used in Yolngu ceremony \citep{norrisbook, Allen75} and in depictions of constellations such as Scorpius in bark paintings (ibid.). What is not well-established  is whether any measurements were ever made  of the positions of the celestial bodies, nor whether there is any ethnographic reference to the solstices or equinoxes.

There are about 400 different Aboriginal cultures in Australia, and it is dangerous to assume similarities between them. On the other hand, it is important to acknowledge that in some cases there are some similarities. For example, the association of Orion with a  young man or group of males, and the association of the Pleiades with a group of girls, are found in many Aboriginal cultures across Australia.  In this paper we focus entirely on the builders of the Wurdi Youang Stone Arrangement, the Wathaurong people, and do not assume any similarities with other Aboriginal cultures, although we refer to them to set context.

\subsection{Stone Arrangements}
Stone arrangements were constructed by several Indigenous cultures across Australia, and include many different morphologies  (e.g. circles, lines, pathways, standing stones, and cairns; \cite{Enright, Towle, Palmer, Lane80, Frankel, Attenbrow}). Some appear to have a practical purpose (e.g. fish traps, land boundaries) and others a  ceremonial purpose (e.g. initiation and burial).  They are often associated with other Aboriginal artefacts such as rock engravings, scarred trees, and axe grinding grooves \citep[e.g.][]{Lane80, lane09}.

Stone arrangements vary in size from a metre to hundreds of metres in length, and are typically constructed from local rocks that are small and could be carried by one or two people, although occasionally they can weigh as much as 500~kg \citep{Lane80, LongSchell}.  Ceremonial stone arrangements are commonly found on ridges and hilltops that command  a panoramic view of the surrounding landscape \citep{Hamacher}.   \cite{McCarthy} suggests that stone arrangements used for ceremonial purposes incorporate the surrounding landscape, and may indicate the direction of a landmark, or mimic a land feature.

While only 30 stone arrangements are recorded in Victoria \citep{marshall}, more are known to the authors, 
and \cite{lane09} claims that hundreds exist in western Victoria.  Unfortunately, no known ethnographic records or oral histories exist about these arrangements, possibly because Aboriginal communities consider sites to be sacred and secret to outsiders \citep{McBryde}.

\subsection{Wurdi Youang}
\begin{figure}[hbt]
\includegraphics[width=8cm]{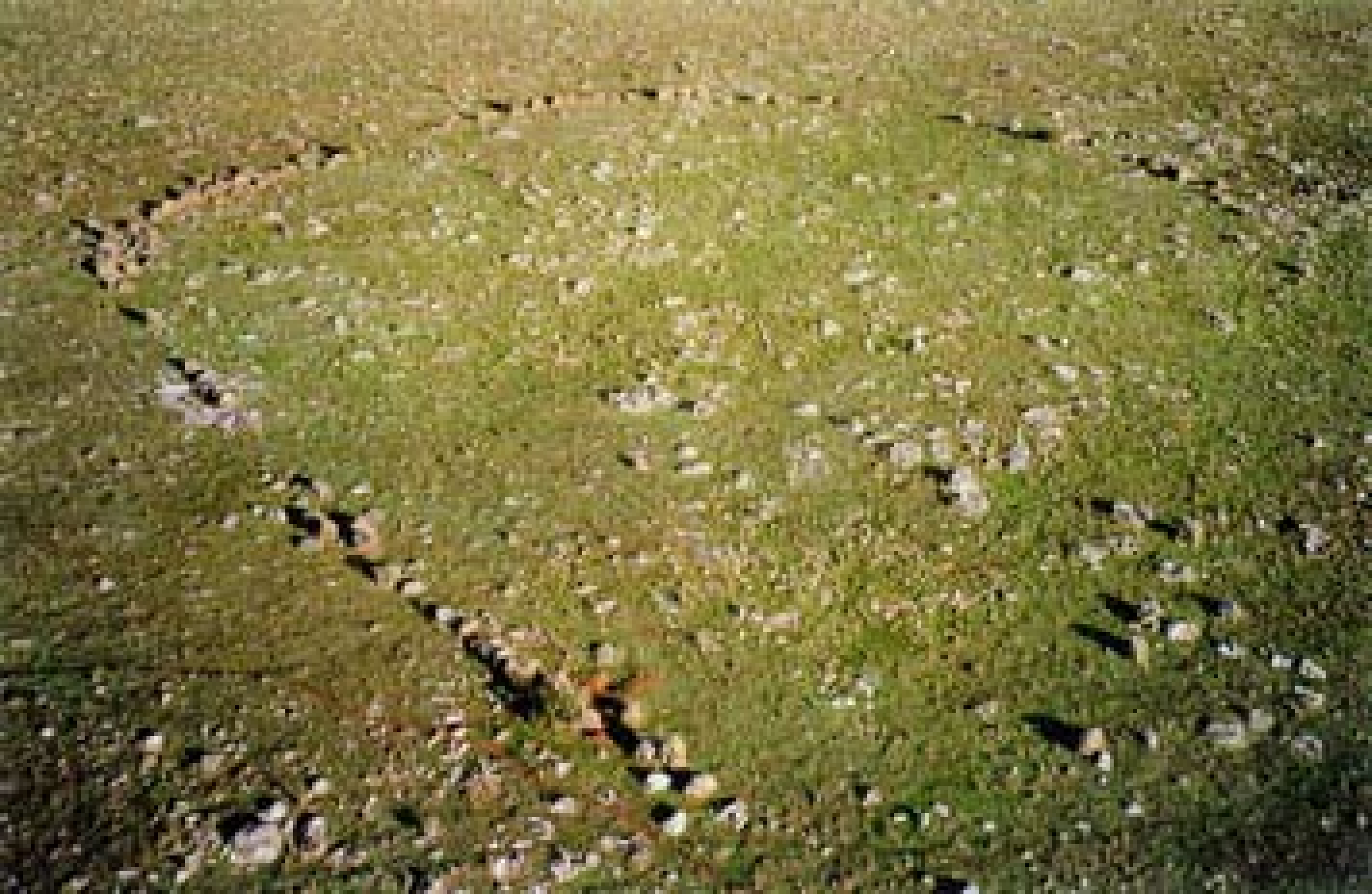}
\caption{Aerial view of the Wurdi Youang site, reproduced with permission from \cite{marshall}, looking west.  } 
\label{aerial}
\end{figure}

The Wurdi Youang stone arrangement, also known as the Mount Rothwell Archaeological Site, is shown in Fig.\ref{aerial}. It lies near the small town of Little River, between Melbourne and Geelong, and was declared a protected site in 1977 by the Victorian Archaeological Survey (AAV Site No. 7922-001). The site is traditionally owned by the Wathaurong people (also known as Wada Wurrung), whose land extends westward from the Werribee River to Fiery Creek beyond Skipton, and northward from the south coast to the watershed of the Great Dividing Range north of Ballarat. To protect the site, the precise location is not given here, but access may be gained after obtaining permission from the traditional owners via Aboriginal Affairs Victoria.  

Wurdi Youang consists of a roughly egg--shaped ring of about 100 basalt stones, about 50~m in diameter along the major axis, which is aligned east--west. The stones range from small rocks about 0.2~m in diameter to standing stones up to 0.75~m high, some of which appear to be supported with trigger stones.  \cite{Lane80} estimate their combined mass to be about 23 tonnes.  They all appear to be potentially movable, rather than being part of the bedrock. 

Particularly prominent are a group of three large stones, about 0.6~m high, at the western end of the stone arrangement. They are located at the highest point of the stone arrangement, which is built on land that slopes downwards from its western end to its eastern end, with a  total fall across the arrangement of about four metres. 
 
 While there exists no eyewitness record of the stone arrangement being constructed or used by the local Wathaurong people, the site is considered Aboriginal in origin for the following reasons \citep{aav}:

\begin{itemize}

\item Similar stone arrangements are known to occur elsewhere in Victoria, although none are known that exactly resemble Wurdi Youang \citep[e.g.][]{massola};

\item The stone arrangement is on a property that has been owned by a single family since first settlement, and the family tradition rules out a European origin \citep{Lane80};

\item The arrangement has no known counterpart among colonial structures: it is on rocky ground with no commercial or agricultural value, it would not have been suitable for defining the boundaries of a sheep dip, sheep pen, or cattle dip, and there is no evidence that it ever formed part of a fence or building \citep{Lane80};

\item The Wathaurong owners have traditional knowledge regarding the sanctity of the site \citep{marshall}.

\end{itemize}

In addition, Aboriginal artefacts have been found on the site by the Wathaurong owners.

Its construction date is unknown.  Aboriginal people are believed to have inhabited the area from about 25 000 BCE \citep{clark1990} to about 1835 when the area was occupied by European settlers \citep{clark}.  It has been suggested \citep{morieson94} that the name ``Wurdi" means ``plenty of people", and Youang means ``bald" or ``mountain", which is presumably related to the nearby mountain range  named the ``You Yangs".  Alternatively, \cite{morieson03} suggested that the name ``Wurdi" may be related to the Woiwurrung word ``Wurding'' meaning abalone, and that the shape of the stone arrangement may be intended to resemble an abalone shell, or possibly another mollusc, in which case it may be conjectured that the site was used for increase rituals. However, these suggestions must be weighed against the distance (18km) of the site from the nearest major body of saltwater (Port Phillip Bay) where abalone could be found.

The vegetation around Wurdi Young is currently low and scrubby, and may have been much higher before European occupation, perhaps even obscuring the view of the setting Sun. However, we also note the common Aboriginal practice of periodically clearing land by fire when necessary,  as part of standard Aboriginal land management practices \citep{Clarke07,Gammage}, so it is equally possible that the vegetation was removed.  If this site was used to observe the position of the setting Sun, then any such growth would have had to be cleared in those directions.

\subsection{The Morieson Hypothesis}

\cite{morieson03}  suggested that three small outlying stones (the ``outliers") marked the setting sun at the solstices and equinoxes when viewed from the three prominent stones at the western apex. Specifically, the Morieson hypothesis is that the outliers were deliberately placed by the Aboriginal builders to indicate the position on the horizon of the setting Sun at the equinoxes and solstices.  The primary purpose of this paper is to test the Morieson hypothesis.

Despite the potential importance of this site to our knowledge of pre--contact Aboriginal cultures, the only two available surveys of this site differed significantly, suggesting that that at least one of them was seriously flawed. Furthermore, neither survey included the outliers proposed by Morieson, and so a new survey was needed to test the Morieson hypothesis.  Since that survey was made, we have become aware of a further survey which is discussed below.

\subsection{Secular Changes in the Sky}
\label{precession}
The rotation axis of the Earth precesses relative to the stars, completing a circle of radius 23.5$\deg$ over a period of about 26000 years. This motion is named the `precession of the equinoxes' and causes the apparent positions of stars to shift by 1$\deg$ every 72 years from the viewpoint of the observer. If the site was used thousands of years ago, the positions of stars would have been significantly different from today. Thus the position on the horizon at which a star sets changes relatively rapidly with time, and a stellar alignment 2000 years ago could differ from that seen today by nearly 30$\deg$.   

Moreover, stars are not stationary, but move relative to their neighbours. This effect, dubbed `stellar proper motion', causes the stars to shift their apparent position relative to each other over time.  For example, the familiar shape of the Southern Cross would have looked significantly different 10000 years ago.

Unlike the positions of stars, the declinations of the Sun and Moon, and hence their rising and setting positions, are unaffected by this precession.  However, the apparent declination of the Sun is affected by a much smaller effect, the nutation in the obliquity of the Earth's rotational axis, which varies by about 2.4$\deg$ over a period of 41000 years.  Because the alignments discussed here are accurate to a few degrees, such variations will have no measurable effect on these alignments. 

Since the construction date has no measurable effect on the rising and setting positions of the Sun, it will not be considered further in this paper.

\subsection{Goals and Outline of this Paper}
This paper adopts an approach taken from the physical sciences in which hypotheses are made and then tested against data. If a prediction of a hypothesis is not confirmed by data, then the hypothesis is rejected. Thus this paper does not attempt to convince the reader that Wurdi Youang is an astronomical site, but merely attempts to test the existing hypothesis that it is. The intention of the paper is  therefore to be impartial, and specifically considers the possibility that any apparent astronomical alignments may be due to pure chance, rather than the intentions of the builders of Wurdi Youang.

In \S\ref{survey} of this paper, we present a new survey of the site, and compare it to previous surveys. In 
\S\ref{morieson}, we use these  data to test the Morieson hypothesis, and in \S\ref{new} we consider other potential alignments. In \S\ref{astronomy} we bring these data together to test whether Wurdi Youang was intentionally constructed to fulfil an astronomical purpose, or whether the alignments are simply due to chance.

\section{Survey and Data Analysis}
\label{survey}

\subsection{The New Survey}
\label{survey:siteplan}

\begin{figure}[hbt]
\includegraphics[width=8cm]{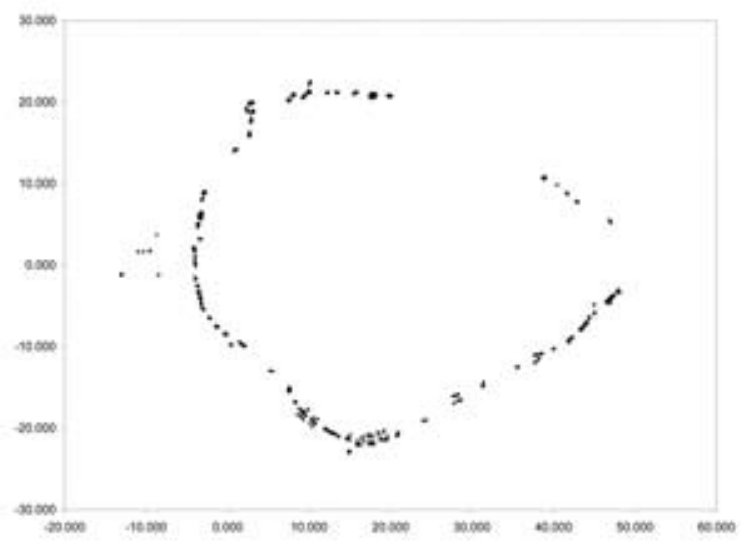}\vspace{2mm}
\caption{Our new survey. Both axes are in metres, and the x-axis is oriented true east. } 
\label{fig2}
\end{figure}

We conducted the main site survey on 8 May 2006. Surveying targets were fixed to prominent stones using Blu-Tac, and the positions of each of these targets was then measured using a Wild--T1 theodolite (accuracy $\sim$1 arcmin) and a laser ranging device (accuracy $\sim$~3~mm). These surveyed positions are shown on Fig. \ref{fig2} as small white circles, and we estimate their positional accuracy to be about 5~mm in all three axes.

The azimuth of the theodolite was calibrated using timed observations of the limb of the Sun, and compared to the position calculated using the ``Horizons" ephemeris software provided by NASA Jet Propulsion Laboratory. Several measurements were made, resulting in a standard error in azimuth of $\sim$~1~arcmin.

A sketch was made by hand of the ground plan of each stone in the ring, using a metre rule and a magnetic compass as guides. The surveyed markers and orientations were shown on these sketches, and we estimate the resulting accuracy to be 10 cm in each horizontal axis. These sketches were then combined with the surveyed positions to yield Fig. \ref{fig2}. As a check, we also photographed each stone from above and from the side with a Canon 350S digital camera. 

A challenge to such surveys is deciding which stones to include in the survey and which to omit.  We tried to include all stones within 5~m of the line of the ring which were greater than 30~cm in size, omitting only those stones which appeared to be part of the bed rock. However, we recognise that there is considerable subjectivity in the choice of stones.  It is also possible that some stones were missed because of the long grass and deteriorating visibility during the survey, although we do not believe that this will greatly influence the overall plan of the site.

A preliminary survey was conducted on 6 May 2006.  The laser ranger was not available, so a surveyors tape measure was used instead. Although a complete survey of the site was made on this day, we have not used the data (which are inferior to the data taken on the 8 May) except as a check for errors on the main survey.
\subsection{Results}
We present the site plan in Fig. \ref{fig2}. Most stones delineate an overall egg-shape, oriented east-west, although there are clearly several apparently irregular deviations from this. At the western end are the five outliers  referred to by Morieson. Three of these are in an approximately east-west line, and a further two are north and south of this line. 

The Morieson hypothesis rests on the position of these small outliers.    Because they are small and easily disturbed, we searched the area for other potential outliers that should be included to avoid biasing the results.  In doing so, we adopted the criteria used throughout the survey, that the stones should exceed 30~cm in one dimension, and should not be bedrock. Only one such outlier (Stone 8, discussed below) was found, other than those noted by Morieson, leading to another potential alignment, discussed below.

\subsection{Comparison with Other Surveys}
\label{survey:comparison}

\begin{figure}[hbt]
\includegraphics[width=8cm]{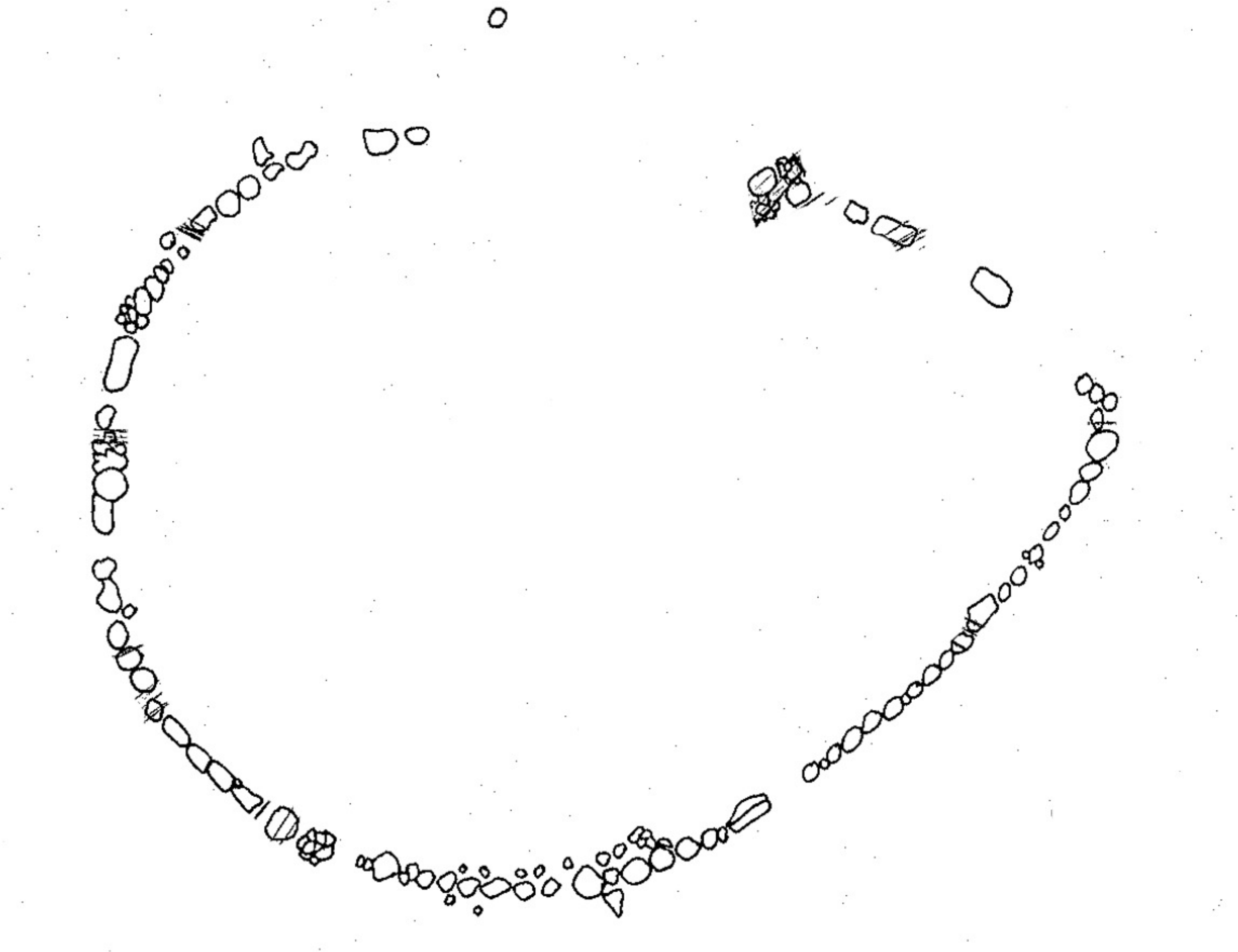}
\caption{The survey published by \cite{Lane80}.}
\label{fig:Lane80}
\end{figure}

\begin{figure}[hbt]
\includegraphics[width=8cm]{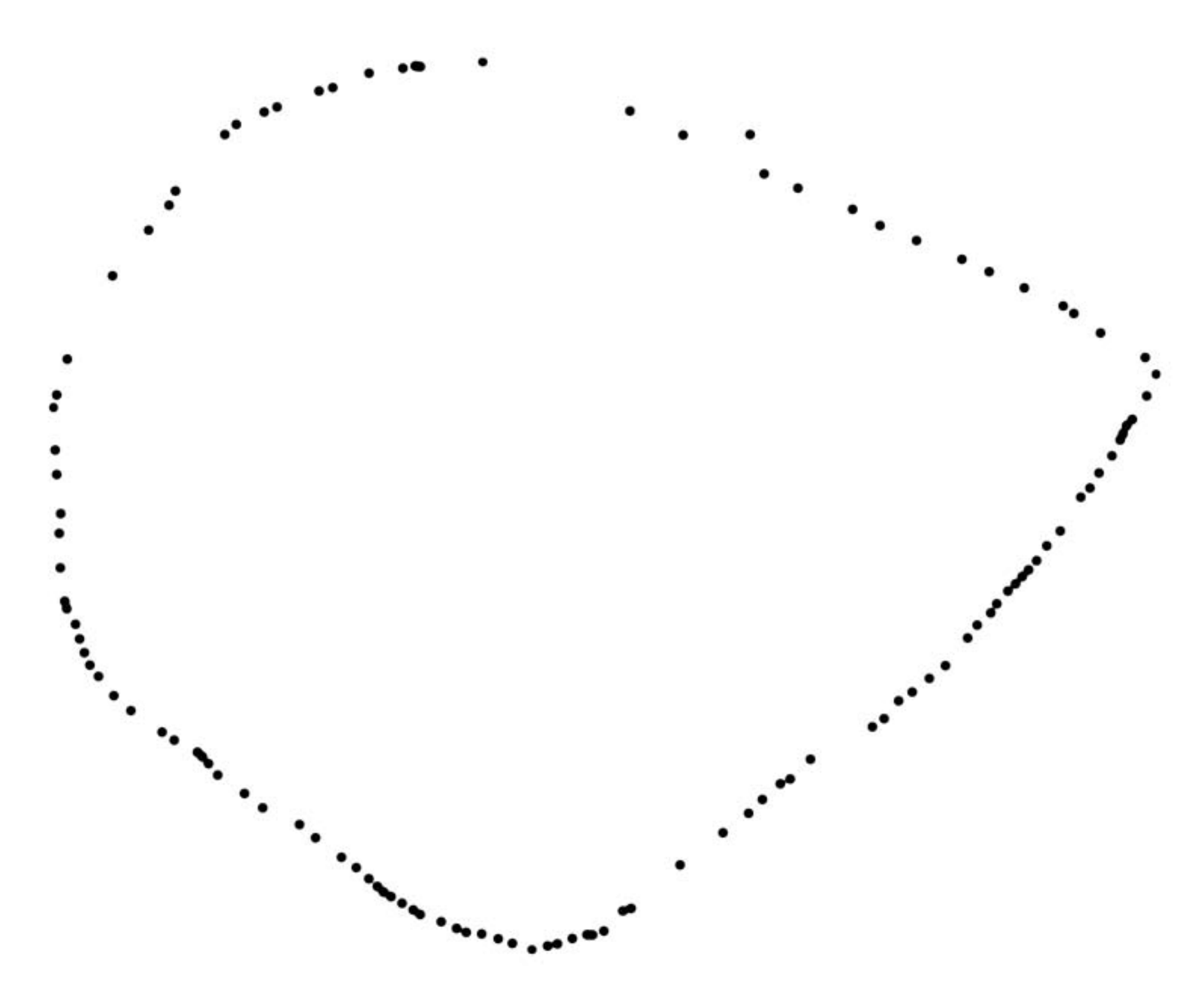}
\caption{A survey obtained by Morieson in about 2000, measured using a tacheometer (theodolite plus laser ranging). Only one position on each stone was measured.} 
\label{fig:M_survey}
\end{figure}

The first written description of Wurdi Youang occurred in 1973--74, in a proposal  cited by \cite{marshall} for the declaration of the site as an archaeological area. The first sketch was by \cite{lane75}, and the first published survey was by \cite{Lane80}, which we subsequently refer to as the Lane \& Fullagar survey, and which is shown in Fig. \ref{fig:Lane80}.  A subsequent survey by \cite{richards96} shows the contours of the land and the location of the site, but does not give a detailed survey of the stone arrangement itself.  A survey (which we refer to here as the Morieson survey, shown in Fig. \ref{fig:M_survey}) with a tacheometer was made in about 2000 by Morieson, but only one point on each stone was recorded, and the shape and orientation of individual stones was not recorded, precluding a full diagram being made. Nevertheless, there was significant disagreement between the Morieson survey and the Lane \& Fullagar survey, suggesting that at least one of them was significantly erroneous.   At the time of our survey, these were the only surveys known to us.

\begin{figure}[hbt]
\includegraphics[width=8cm]{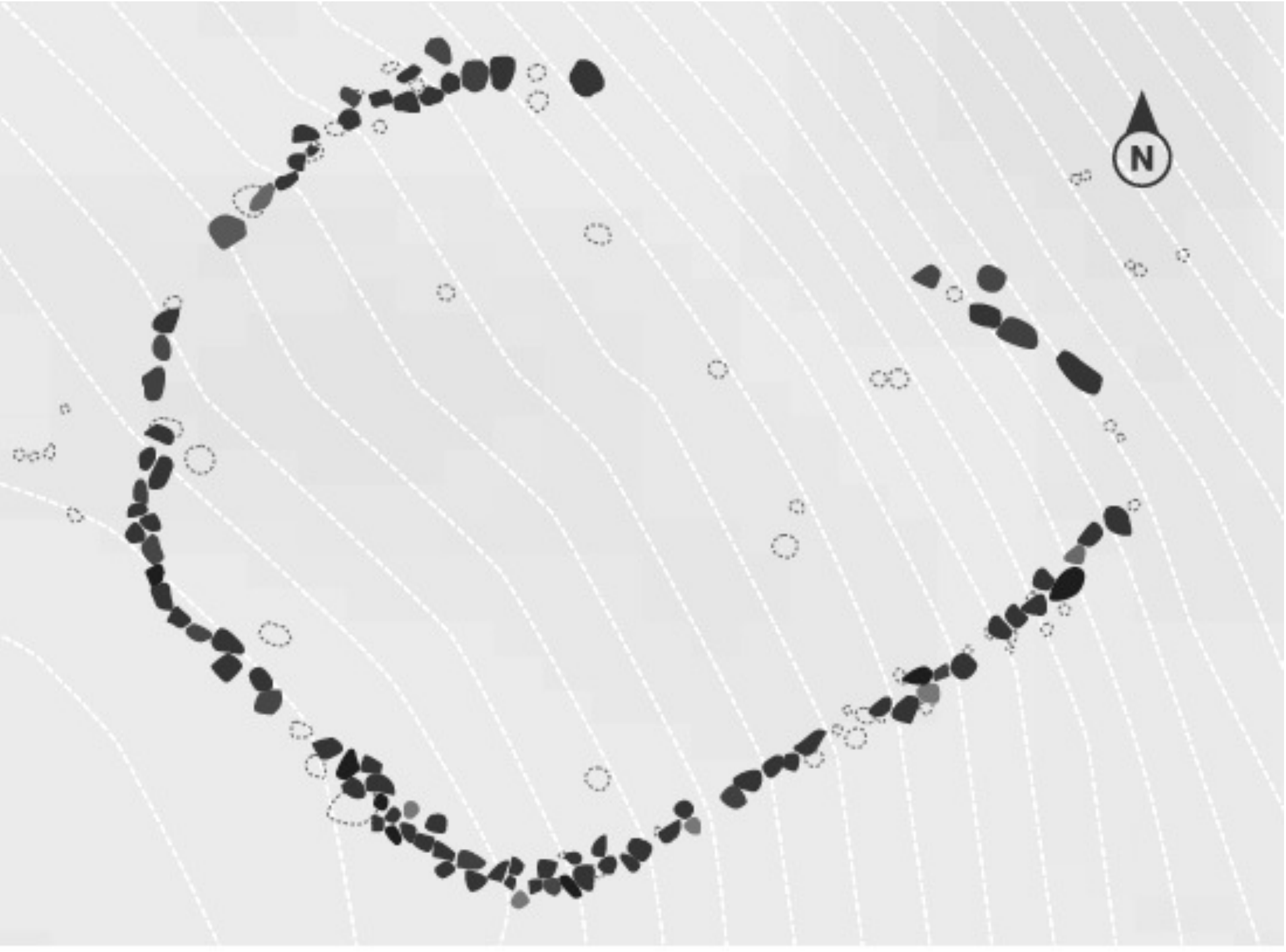}
\caption{The survey made by \cite{marshall} for Aboriginal Affairs Victoria, reproduced with permission.}
\label{fig:AAV}
\end{figure}

After we had conducted our survey, we became aware of a further  accurate survey reported by \cite{marshall}, which was commissioned by Aboriginal Affairs Victoria, and conducted in 1998, and which we refer to here as the `AAV survey', shown in Fig. \ref{fig:AAV}.  A comparison shows substantial agreement  between our survey, the Morieson survey, and the AAV survey.  The enlarged section of the critical westernmost section, shown in Fig. 6, illustrates the similarities and differences between the surveys.

\begin{figure}[hbt]
\includegraphics[width=8cm]{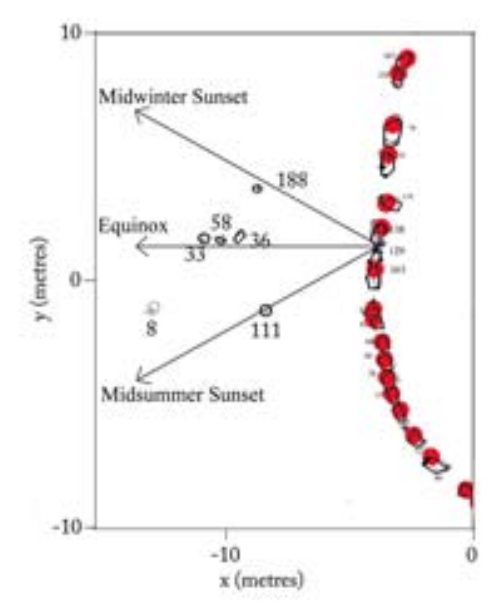}
\caption{An enlargement showing the three surveys. The bold outlines are from the AAV survey, the fainter lines (which in many places are coincident with the bold lines) are from our survey, and the red dots are from the Morieson survey. Crosses show the locations of surveyed points in our survey, and numbers are the reference numbers of each stone used in our survey.} 
\label{detail}
\end{figure}

Of the four surveys, the greatest agreement was found between our survey and the AAV survey.  The overall shape of the arrangement is similar and the positions of nearly all individual stones agree to within the stated uncertainties.  The main difference between these surveys is the choice of stones selected as part of the overall arrangement.  This is to be expected, as it is a subjective decision whether or not to include a particular stone.  Those stones which are common to the two surveys typically agree in position and outline to within $\sim$~10~cm. 

An apparent difference between our survey and the AAV survey arises because the AAV survey was measured relative to the Australian Map Grid North, whereas ours was measured relative to True North (as derived from astronomical observations). The AAV survey must therefore be rotated counter--clockwise by 1.542$\deg$ (calculated using the Redfearn tool on the Geoscience Australia website\footnote{http://www.ga.gov.au/earth-monitoring/geodesy/geodetic-datums.html}) before being superimposed on our survey.

The electronic survey by Morieson agreed with our survey and the AAV survey in terms of overall shape.  However, only a rough comparison may be made between these surveys since the Morieson survey recorded only the relative location of single points on the stones, and did not record either the overall bearing relative to north, nor the shape of each stone. Nevertheless, it is in substantial agreement with both surveys.

In contrast, the survey by \cite{Lane80} is very different from the three other surveys, despite their statement that ``Sites were surveyed with a theodolite. The above--ground dimensions of all stones were measured...''.  The shape of the arrangement, in their plan, is a smooth egg-shape, unlike the irregular shape shown in the other three surveys, and the individual stones in their plan do not appear to resemble those shown in the other surveys. We conclude that the statement in \cite{Lane80} quoted above is erroneous, and that their Fig. 32 is in fact a rough hand--drawn sketch and is not based on measurements of the positions of individual stones.  It will not be discussed further in this paper.

The Morieson hypothesis rests on the existence of the small outlying stones to the west of the ring.  These are not shown in the \cite{Lane80} survey nor in the Morieson survey, but appear both in the AAV survey and in our survey.  In our survey, we adopted the criteria that the stones should exceed 30~cm in one dimension, and should not be bedrock. All five Morieson outliers satisfy these criteria, as did a further  outlier (Stone 8), which was not noted by Morieson, but which generates another potential alignment, discussed below.

\begin{figure}[hbt]
\includegraphics[width=8cm]{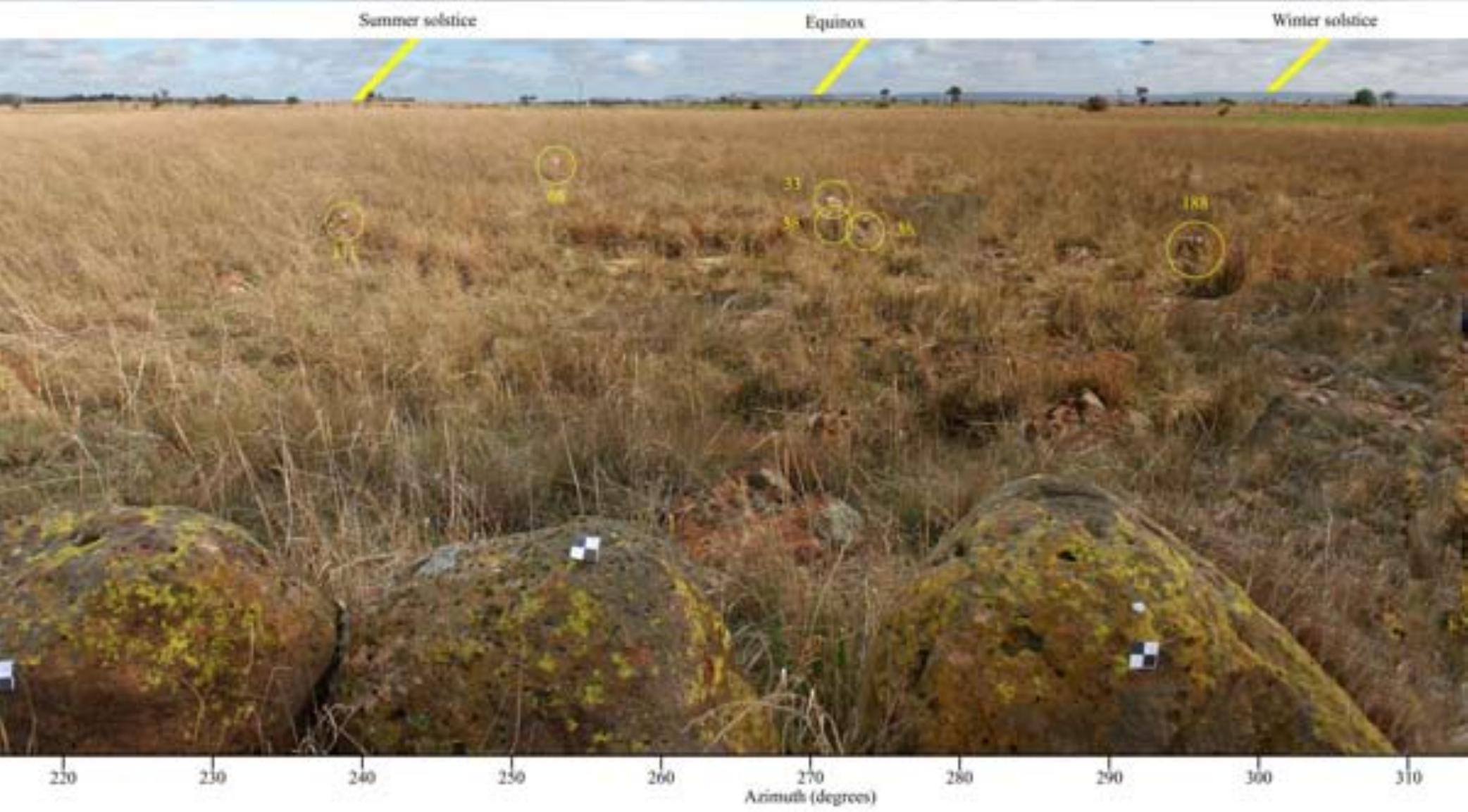}
\vspace{2mm}
\caption{The horizon above the Morieson  alignments, assembled from a mosaic of individual photographs. Azimuths are in degrees relative to True North. The setting positions of the Sun at the equinoxes and solstices are calculated using the JPL Horizons software. In the foreground can be seen three of the surveying markers used during our survey, affixed to the major stones at the western apex. The outliers are indicated by yellow circles. Numbers are those used in our survey, also shown in Fig. 6. The apparent misalignment of stones 33, 58 and 36 is caused by foreshortening in the image - their correct positions are shown in Fig. 6.} 
\label{panorama}
\end{figure}

\section{Testing the Morieson\\ Hypothesis}
\label{morieson}

In this section, we test whether our survey supports the Morieson hypothesis, in four stages. 
\begin{enumerate}
\item We test whether the Morieson hypothesis is consistent with the survey, in that the alignments suggested by Morieson are actually correct. 
\item Given the stones at Wurdi Youang as they are now, we determine what other alignments are as likely as the Morieson alignments 
\item We test whether, if we were to build many copies of Wurdi Youang with stones randomly placed in positions similar to the current Wurdi Youang, a significant number of these would have as many significant astronomical alignments as the real Wurdi Youang (i.e. whether the alignments could  merely be chance alignments and not intentioned by builders). 
\item We note that the overall shape of the stone arrangement also indicates potentially astronomical directions, and we estimate the likelihood of this occurring by chance.
\end{enumerate}

The Morieson hypothesis is illustrated by Fig. \ref{panorama}. Morieson's suggested viewing position is from a gap (hereafter ``the Gap'') between two of the three large rocks at the western apex of the arrangement. He then suggests that the lines of sight over the outliers from this viewing position indicate places on the horizon where the sun sets at the two equinoxes and solstices.



\subsection{Accuracy of the Morieson\\ Hypothesis}
\label{morieson:accuracy}

Morieson framed his hypothesis by observing the setting sun on the relevant days.  Here, we test whether the survey supports his observation that, when viewed from the Gap, the Sun sets above the outliers on the solstices and equinoxes. 

From our survey, and by using the ``Horizons" online tool provided by NASA/Caltech Jet Propulsion Laboratory, we calculate the position of the Sun at the solstices and equinox. The result is listed in Table \ref{tab1} and shown in Fig. 7.

Both the equinoctial alignment and the summer solstitial alignment are correct to within $\sim$~2$\deg$ and the winter solstitial alignment is accurate to $\sim$~3$\deg$. This may be compared with the width of the outliers as viewed from the large stones ($\sim$~2$\deg$) and the change in orientation if the viewer moves his/her head sideways by one head--width ($\sim$~2$\deg$).  We conclude that the positions of the stones are consistent with the Morieson hypothesis. 

\subsection{Significance of the\\ Morieson alignments}  
\label{morieson:stats}

Given all the stones at Wurdi Youang, many lines of sight may be drawn between them, and, purely by chance, some may happen to indicate the position on the horizon of an astronomically significant event.  If a researcher selectively reports only those astronomically significant alignments,  an astronomically significant site may be spuriously created.  Here we test whether this has occurred in the case of the Morieson hypothesis. To accomplish this, we make no assumption about viewing position, but estimate how many other alignments could be identified by an unbiased researcher, that are indicated by the outliers, and that are at least as significant as those identified by Morieson.  

To make the question tractable, we need to make some reasonable assumptions to limit the range of potential viewing positions. For example, thousands of lines might be drawn between individual stones around the perimeter of the arrangement, but of course these are far less ``obvious'' than lines between particularly prominent stones and the outliers. We therefore restrict ourselves to lines indicated by the outliers, and use the  term ``prominent'' throughout the rest of this paper to refer to such distinctions. For example, a ``prominent'' viewing position is one that we consider any reasonable unbiased observer would choose (e.g. a large stone, or the top of a hill) over less prominent viewing positions (at any one of a miriad of small stones, or half way up a hill).

In principle, it is possible that Morieson selected only those outliers that suited his hypothesis, and so we searched for other potential outliers which he may not have recorded, but which satisfied our criteria (not bedrock, and greater than 30~cm in one dimension).  Only one such additional outlier was found (Stone 8) and is included as a potential outlier for our statistical analysis.

We consider potential viewing positions from which the outliers may be seen. Prominent positions are:
\begin{enumerate}
\item the centre of the stone arrangement 
\item the crest of the hill (which coincides with the three large stones at the western apex identified in the Morieson hypothesis).  This may be subdivided into
\begin{enumerate}
\item behind the centre of each of the large stones
\item from the gap between each of the large stones
\end{enumerate}
\end{enumerate}
However, the outliers are not visible from the centre of the ring, leaving only the five viewing positions (i.e. 3 stones and two gaps) in the vicinity of the three large stones.    Furthermore, the two southernmost stones (Stones 163 and 129 in Fig. 6) have no gap between them, leaving just four potential viewing positions (i.e. 3 stones and the single Gap).


As a result, there are four potential viewing positions, each of which has four potential alignments, listed in Table \ref{tab1} .  If we regard an arbitrary limit of 5$\deg$ as the limit of a correct alignment, then the Morieson viewing position has three astronomical alignments, and the other viewing positions have two, two, and zero respectively, confirming that the Morieson viewing position is the best position for viewing the astronomical alignments. Nevertheless, in the next step of the analysis, we ignore this and consider all 16 alignments as equally likely. 

The occurrence of astronomically significant alignments from these other viewing positions is unsurprising, as the viewing points are close together. For example, stone 111 indicates the position of the summer solstice within 5 $\deg$ regardless of whether its is observed from the Gap or from either of the adjacent stones.

\subsection{Monte Carlo analysis}  
\label{mc}
We now estimate the likelihood that the sixteen alignments have occurred by chance. We do so by employing a Monte Carlo simulation, in which we simulate the building of ten thousand stone arrangements, each similar to Wurdi Youang, but moving the viewing positions and the outliers randomly, while keeping them in the same general area of the stone arrangement. We then introduce a ``scoring" system to measure how well the new stone arrangement indicates the astronomical setting positions.

We approximate the four viewing points by distributing them randomly in a line 2~m long, and approximate the four outliers by distributing them randomly along a line 5~m long, located 4~m from the first. We then consider all the alignments from the viewing positions which pass over the outliers and intersect the horizon. 

Our scoring system is necessarily arbitrary, but hopefully unbiased. If an alignment points within 2$\deg$ of the setting sun at the solstices or equinoxes, it is considered a `hit' and given a score of 1.  Near misses are given lower weights which reduces linearly with distance from the hit, with a miss of 4$\deg$ having a score of 0.5, and 8$\deg$ a score of 0.25. For each simulated arrangement, we then take the sum of these scores for three stones indicating the equinox and the solstices.  Applying this algorithm to our surveyed Wurdi Youang data gives it a total score of 2.565. 

The simulation was run 10000 times (i.e. we simulated the building of Wurdi Youang 10000 times), 25 of which had a score $>$2.565, which implies that the likelihood of the Wurdi Youang alignments occurring by chance is 0.25\%. We recognise that this is not a precise calculation, and that our simple approximation to the geometry of the site introduces the possibility of bias, but nevertheless this process provides a rough estimate of the likelihood of the alignments being due to chance alone.

We conclude that it is extremely unlikely that the outlier stones happen to indicate the astronomical alignments by chance. Instead, we conclude that the alignments were almost certainly constructed deliberately by a human being to indicate the equinoxes and solstices.

\subsection{Newly Identified Alignments}
\label{new}

As well as the alignments of the Morieson hypothesis, another prominent alignment is the major axis of the stone arrangement, which lies roughly on an east--west axis.  A prominent viewing position for this would be either the centre of the stone arrangement, or the eastern apex (the lowest part of the arrangement). In practice, because of the local terrain, both these viewing points indicate roughly the same place on the horizon, which is due west of the site, or the position at which the equinoctial Sun sets.  

It is important to consider whether there are any other  prominent alignments at the site. 
The egg--shape of the stone ring includes two nearly straight sections to the east of the ring, clearly visible in Fig. 1, which also constitute prominent alignments. We consider that these would be chosen as the only prominent alignments in the ring of stones by an unbiased observer, whether looking at the plan or being physically present at the site.
In each case we consider only the western--facing direction of the alignment, since the site is built on a slope that rises to the west.  and so the westerly lines point to the horizon, whilst the easterly counterparts point down into the valley. While Fig. 1 indicates that there \emph{is} a relatively straight section in each case, the precise direction is poorly defined, both because it depends on the choice of stones to be included (in a  least-squares fit, for example), and because of damage to the stones. However, the direction of each straight section is roughly parallel to the Morieson alignments, as shown in Fig. 8, where the directions of the equinox and solstices are shown superimposed on both the ring and on the Morieson alignments. 

From the eastern vertex of the stone arrangement, as defined by the intersection of the straight lines in Fig. 8, the Gap  is at an azimuth of 272$\deg$ and an elevation (from a viewing height of 1.6m) of 2$\deg$. Thus, viewed from the Vertex, the setting equinoctial Sun would set directly behind the three prominent stones at the western apex of the arrangement, and, depending on the exact position and height of the viewer, would be briefly visible through the Gap before setting.

\begin{figure}[hbt]
\includegraphics[width=8cm]{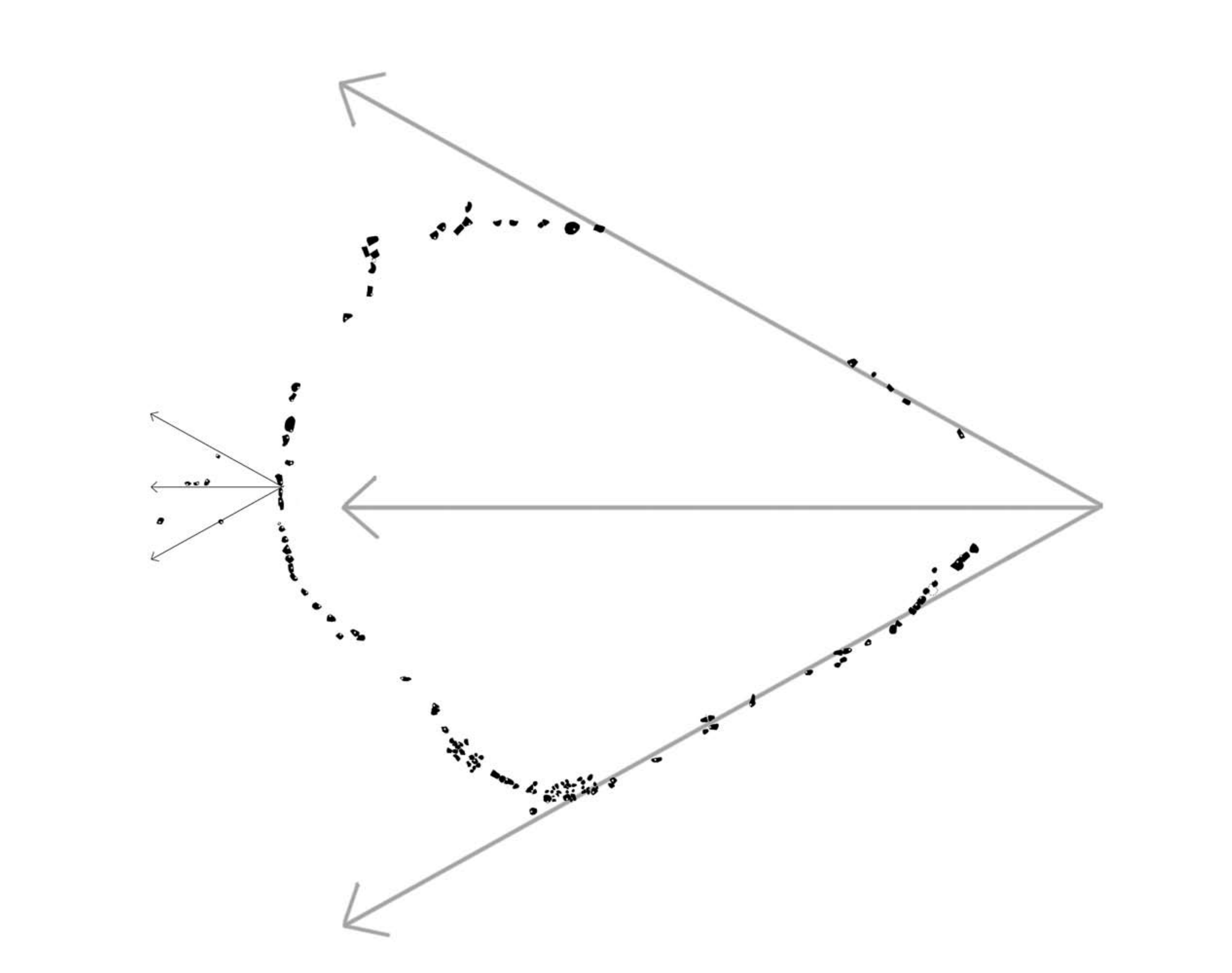}
\vspace{2mm}
\caption{The arrows indicating the directions to the equinoxes and solstices superimposed on the the outliers, left, and the ring (right). Note that these directions are \emph{not} adjusted to fit the ring, but are defined astronomically. Thus while the straight sections of the ring are not well defined, and not exactly straight, this diagram shows that they are well aligned to the same astronomical directions as the Morieson alignments.} 
\label{alignments}
\end{figure}

If we include the outliers as prominent alignments, and choose the viewing position to be that chosen by Morieson, we consider that there are a total of seven prominent alignments: 
\begin{itemize}
\item the three noted by Morieson,
\item a fourth over the new outlier, 
\item the major axis of the ring, and 
\item the two almost straight sections of the ring. 
\end{itemize}

We summarise these alignments in Table \ref{tab2}.

For each potential alignment, the azimuth and elevation of features on the indicated horizon were measured using the theodolite, and the horizon was also photographed using the digital camera. The photographs were then combined in Adobe Photoshop, using the surveyed horizon positions to align them, resulting in the panoramic images shown in Fig. 7.  No correction was made for lens distortion, but photographs were taken centred on the horizon above each outlier to minimise the distortion there.

Also shown on Fig. 7 are the paths of the setting Sun at the solstices and equinoxes in 1788 CE, calculated using the Horizons software.  We do not have archaeological information to date this site, so this date was chosen as approximating the last date on which this arrangement might have been used by the Wathaurong people before their culture was damaged by the influx of European settlers. However, as described in \S \ref{precession}, solar positions change very slowly with time.  4000 years ago, the positions of the Sun at the solstices would be different by only 0.5$\deg$ in azimuth from those shown here, and the position of the Sun at the equinox would be identical.

Of the seven potential alignments shown in Table \ref{tab1}, six are astronomically significant to within a few degrees.  It is difficult to use a Monte Carlo analysis to estimate a likelihood of this arising through chance, because it is almost impossible to estimate the likelihood that two straight lines occur by chance in the egg-shaped arrangement. Nevertheless, the likelihood that these are chance alignments is extremely small (enormously less than than the 0.25\% estimated for the Morieson alignments alone).  

Given that three of the four indicated alignments over the outliers are astronomical, it is possible that the direction over stone 8 may also be astronomical. This direction does not correspond to any significant solar or lunar alignment, but could align with the setting position of bright stars. However, the rapid precession of the stars means that almost any azimuth will have indicated some significant  stellar setting place at some point in the last 50,000 years. It is therefore impossible to determine whether any such alignment, taken in isolation, is significant.

\section{Is Wurdi Youang Astronomical?}
\label{astronomy}

We have shown above that the Morieson hypothesis is supported, in that the azimuths over the outliers do indeed indicate the position of the setting Sun on the equinoxes and solstices, and that the likelihood of this occurring by chance is low (about 0.25\%).  We have also shown that three other alignments are also astronomical in nature.  The likelihood of this occurring by chance is very close to zero, suggesting that these alignments are almost certainly intentional. We caution against over--interpreting these calculated probabilities, as they do not take into account our potential cultural bias. Nevertheless, they strongly suggest that the stone arrangement was deliberately intended by its builders to point to the setting Sun at the solstices and equinox.

The larger stones at Wurdi Young, firmly supported by trigger stones, would be unlikely to move under the effects of soil erosion. However, the smaller stones, particularly the outliers, are easily movable by hand, and it's important to consider whether they may have been moved by humans or animals.  If they have been moved, the likelihood of their coming to rest in these astronomically-significant positions by chance is negligible. To have been moved into these positions would require a deliberate hoax prior to the AAV survey in 1998, a hypothesis for which there is no supporting evidence. Furthermore, the newly identified lines in the structure of the stone arrangement argues against a hoax.

Although we find that the site was deliberately constructed to indicate significant solar positions, the primary purpose of this arrangement may not be astronomical in nature and we are careful not to label this an ``Aboriginal observatory".  The solar alignments may have been ancillary to the primary function of the site (e.g. ceremonial).  We also find no special alignments between the three stones at the western apex and the three mountains in the background that \cite{morieson03} has suggested they mimic.  Their similarity may be coincidental or have a meaning that is unknown to us.


Finally, despite the apparent alignments at this site, there are no other known sites that share similar solar alignments in Victoria.  While other stone arrangements in southeastern Australia appear to align to cardinal points \citep{morieson94, norris11, Hamacher}, it is currently unknown whether any of these sites have stone arrangements that align to the solar points.  Such a study could lend support to the astronomical hypothesis at Wurdi Youang.  Research into astronomical alignments of stone arrangement sites in Victoria and New South Wales is continuing  \citep{Hamacher}.

\section{Conclusion}
\label{conclusion}

Our detailed survey of Wurdi Youang supports the Morieson hypothesis that a series of outlier stones marks the position of the setting sun at the solstices and equinoxes.  A statistical analysis shows that the likelihood of this occurring by chance is extremely low.  Additionally, we find that the straight sides of the arrangement also indicate the solstices while the three prominent stones at the western apex of the arrangement, as viewed from the eastern apex, mark the point where the sun sets at equinox.  

It is important to note that in this paper we make no {\it assumption} about viewing position, but are testing a specific {\it hypothesis} that has been made about viewing position. Many other sight-lines are possible, and we have tested these using a Monte-Carlo analysis. The analysis shows that the viewing position and the orientations suggested by Morieson are significant, and are  unlikely to have arisen by chance. 

While we do not know the age or purpose of the stone arrangement, we can say with relative confidence that these alignments were intentional, though we are careful not to claim this as an ``Aboriginal observatory" as there exist no known ethnographic or oral histories about the purpose or use of the site.  Further research is planned both to determine the construction date of the arrangement, and to locate any similar sites elsewhere.

\section{Acknowledgements}
We acknowledge and pay our respects to the traditional owners and elders, both past and present,  of the Wathaurong people. We thank the Wathaurong Aboriginal Cooperative for enabling our access to this site and for permitting this research to be conducted.  We thank John Morieson for his advice and guidance, and for initially showing us round the site.  We also thank Cathy Webb and Brendan Marshall of Terraculture Pty Ltd. for trusting us with their last copy of the Webb \& Marshall report, and for permission to reproduce parts of it here, and we thank Brad Duncan of Aboriginal Affairs Victoria for his advice.  We thank the staff of the NASA/Caltech Jet Propulsion Laboratory for freely providing the ``Horizons" tool as an invaluable service to the community, and Barry Parsons for facilitating the loan of the theodolite.\\

\clearpage
\setcounter{figure}{0}
\onecolumn

\begin{table}[h]
\begin{center}
\caption{\emph{Potential} Alignments including the outliers at the Wurdi Youang site.  `Gap' refers to the gap between stones 129 and 138, proposed as the viewing position by \cite{morieson03}.  Az and El are the Azimuth (relative to True North) and Elevation (to the Horizon, from a viewing position 1.6 m above the ground) of the alignment, and Dec is the J2000 Declination of an astronomical object setting at that position, including the effects of refraction.The nominal declinations of the Sun at the winter solstice, equinox, and summer solstice are -23.4$\deg$, 0$\deg$, and +23.4$\deg$ respectively, and a horizon elevation of 0.8$\deg$ is assumed throughout. The ``offset'' column shows the distance from the indicated alignment to the nearest significant astronomical position. If this distance is less than 5$\deg$, the name of the astronomical position is given in the final column.}
\label{tab1}
\vspace{1cm}
\begin{tabular}{lllll}
\hline
Alignment		&	Az$^{\circ}$  	& Offset$^{\circ}$		& 	Astronomical\\
 			& 					& 				&	Significance\\
\hline
Gap to stone 188 & 	296.7 & 4.5 	&	Winter solstice  \\
Gap to stones 36,58,33 & 272.8  &2.1 &Equinox  \\
Gap to stone 8 & 254.3  &13.6 &None  \\
Gap to stone 111 & 240.3  &0.4 &Summer solstice  \\

stone 138 to stone 188 & 	291.4 &	9.8 &	 None  \\
stone 138 to stones 36,58,33 & 268.5  &2.1 &Equinox  \\
stone 138 to stone 8 & 251.4  &10.7 &None  \\
stone 138 to stone 111 & 235.8  &4.9 &Summer solstice  \\

stone 129 to stone 188 & 	301.9 &	0.7 	&	Winter solstice  \\
stone 129 to stones 36,58,33 & 276.9  &6.2 &None  \\
stone 129 to stone 8 & 257.1  &13.6 &None  \\
stone 129 to stone 111 & 244.8  &4.1 &Summer solstice  \\

stone 163 to stone 188 & 	308.8 &	7.6 	&	None  \\
stone 163 to stones 36,58,33 & 293.4  &12.7 &None  \\
stone 163 to stone 8 & 261.8  &8.8 &None  \\
stone 163 to stone 111 & 253.5  &12.9 &None  \\

\hline
\end{tabular}
\end{center}
\end{table}

\begin{table}[h]
\begin{center}
\caption{Significant Indicated Alignments at the Wurdi Youang site. Nomenclature and assumptions are the same as in Table \ref{tab1}.}
\label{tab2}
\vspace{1cm}
\begin{tabular}{lllll}
\hline
Alignment		&	Az$^{\circ}$ 	& El$^{\circ}$ 	& Dec$^{\circ}$		& 	Astronomical\\
 			& 				& 			& 				&	Significance\\
\hline
Gap to stone 188 & 	296.7 			& 0.8 		&	20.5 			&	Winter solstice  \\
Gap to stones 36,58,33 & 272.3 & 0.8 &1.6 &Equinox  \\
Gap to stone 8 & 254.3 & 0.8 &-12.6 &None  \\
Gap to stone 111 & 240.3 & 0.8 &-23.3 &Summer solstice  \\
Vertex to Gap & 272.0 & 2.0 &0.5 &Equinox  \\
line on S side of arrangement & 239.0 & 1.0 &-24.4 &Summer solstice  \\
line on N side of arrangement & 304.0 & 0.8 &25.9 &Winter solstice  \\
\hline
\end{tabular}
\end{center}
\end{table}

\clearpage

\clearpage

\bibliographystyle{aj}
\bibliography{paper_refs}

\begin{thebibliography}{99}

\bibitem[Aboriginal Affairs Victoria, 2003]{aav}Aboriginal Affairs Victoria, 2003, {\it Aboriginal Stone Arrangements}.  Melbourne: Aboriginal Affairs Victoria.

\bibitem[Allen, 1975]{Allen75}Allen, Louis A, 1975, {\it Time before morning}, New York: Cowell

\bibitem[Attenbrow, 2002]{Attenbrow} Attenbrow, V., 2002, {\it Sydney's Aboriginal Past}.  Sydney: University of New South Wales Press.

\bibitem[Cairns \& Harney, 2003]{harney}Cairns, H., \& Yidumduma Harney, B., 2003, {\it Dark Sparklers: Yidumduma's Aboriginal Astronomy}.  Sydney: Privately Published by Hugh Cairns. 

\bibitem[Clark, 1995]{clark} Clark, Ian D., 1995, {\it Scars on the Landscape. A Register of Massacre sites in Western Victoria 1803-1859}.  Canberra: Aboriginal Studies Press. 

\bibitem[Clark, 1990]{clark1990} Clark, I.D., 1990, {\it Aboriginal Languages and Clans: An historical atlas of western and central Victoria, 1800--1900}.   Melbourne: Department of Geography \& Environmental Science, Monash University.

\bibitem[Clark, 1991]{clark91} Clark, A., 1991, {\it Lake Condah Project, Aboriginal Archaeology -- Resource Inventory}.  Melbourne: Department of Conservation and Environment. Victoria Archaeological Survey, Occasional Report 36.

\bibitem[Clark, 1994]{clark94} Clark, A., 1994,  Romancing the stones. The cultural construction of archaeological landscape in the western district of Victoria,  {\it Archaeology in Oceania}, Vol. 29, pp. 1--15.

\bibitem[Clarke, 1997]{clarke97} Clarke, P.A., 1997, The Aboriginal Cosmic landscape of Southern South Australia,  {\it Records of the South Australian Museum}, Vol. 29, No. 2, pp. 125--145.

\bibitem[Clarke, 2007]{Clarke07} Clarke, P.A., 2007,  {\it Aboriginal people and their plants},  Rosenberg Publishing Pty, Ltd., NSW.

\bibitem[Coutts, 1978]{Coutts} Coutts, P.J.F., Frank, R.K. \& Hughes, P., 1978, {\it Aboriginal engineers of the western district, Victoria}.  Melbourne: Ministry for Conservation. Records of the Victorian Archaeological Survey, Vol. 7.

\bibitem[Enright, 1937]{Enright} Enright, W.J., 1937, Notes on the Aborigines of the North Coast of New South Wales.  {\it Mankind}, Vol. 2, pp. 88--91.

\bibitem[Frankel, 1982]{Frankel} Frankel, D., 1982, Earth rings at Sunbury, Victoria Archaeology.  {\it Oceania}, Vol. 17, pp. 83--89.

\bibitem[Gammage, 2011]{Gammage} Gammage, B., 2011, {\it The Biggest Estate on Earth}, Sydney: Allen \& Unwin

\bibitem[Hamacher et al, 2012]{Hamacher} Hamacher, D.W., Fuller, R.S. \& Norris, R.P., 2012, Orientations of linear stone arrangements in New South Wales.  {\it Australian Archaeology}, in press.

\bibitem[Hamacher \& Norris, 2011]{hamachernorris2011} Hamacher, D.W. \& Norris, R.P., 2011, Eclipses in Australian Aboriginal Astronomy. {\it Journal of Astronomical History and Heritage}, Vol. 14, No. 2, pp. in press

\bibitem[Hamacher \& Norris, 2011]{HamacherISAAC} Hamacher, D.W. \& Norris, R.P., 2011, Bridging the Gap through Australian Cultural Astronomy. {\it Archaeoastronomy \& Ethnoastronomy: building bridges between cultures}, Cambridge University Press, ed. Clive Ruggles, pp. 282--290.

\bibitem[Haynes, 1992]{haynes}Haynes, R.D., 1992, Aboriginal Astronomy.  {\it Australian Journal of Astronomy}, Vol. 4, pp. 127--140.

\bibitem[Johnson, 1998]{johnson}Johnson, D., 1998, {\it Night skies of Aboriginal Australia: a noctuary}.  Sydney: University of Sydney Press, Oceania Monographs, No. 47.

\bibitem[Lane \& Fullager, 1980] {Lane80} Lane, L.,  \& Fullagar, R., 1980, Previously unrecorded aboriginal stone arrangements in Victoria.  {\it Records of the Victorian Archaeological Survey}, No. 10 (June), pp. 134--151. Ministry for Conservation, Victoria.

\bibitem[Lane, 1976] {lane75} Lane, L., 1976, The Magic Circle, {\it Geelong Naturalist}, May 1976, 1-6.

\bibitem[Lane, 2009] {lane09} Lane, L., 2009, {\it Aboriginal stone structures in southwestern Victoria}.  A report to Aboriginal Affairs Victoria.

\bibitem[Long \& Schell, 1999]{LongSchell} Long, A. \& Schell, P., 1999, {\it Lake Bolac stone arrangement (AAV 7422-394); management plan}.  An unpublished report to Aboriginal Affairs Victoria.

\bibitem[Marshall \& Webb, 1999]{marshall} Marshall, B., \& Webb, C., 1999, {\it Mount Rothwell Archaeological Area Cultural Heritage Management Plan}.  Report commissioned by Aboriginal Affairs Victoria.

\bibitem[Massola, 1963]{massola}Massola, A., 1963, Native Stone Arrangment at Carisbrook.  {\it The Victorian Naturalist}, Vol. 80, pp. 177--180.

\bibitem[McBryde, 1974]{McBryde} McBryde, I. 1974, {\it Aboriginal Prehistory in New England: An Archaeological Survey of Northeastern New South Wales}.  Sydney: University of Sydney Press.

\bibitem[McCarthy, 1940]{McCarthy} McCarthy, F.D., 1940, Aboriginal stone arrangements in Australia.  {\it The Australian Museum Magazine}, pp 184--189.

\bibitem[Morieson, 2003]{morieson03} Morieson, J., 2003, {\it Solar-based Lithic Design in Victoria, Australia}.  Unpublished manuscript presented at ``Archaeology of the Old World and Oceania, World Archaeological Congress'', Washington D.C., June, 2003.

\bibitem[Morieson, 1994]{morieson94} Morieson, J., 1994, {\it Aboriginal Stone Arrangements in Victoria}.  Unpublished essay, Australian Centre, University of Melbourne.

\bibitem[Mountford, 1956]{mountford} Mountford, C. P., 1956, {\it Records of the American-Australian Scientific Expedition to Arnhem land. Volume 1: Art, Myth and Symbolism}. Melbourne, Melbourne University Press.

\bibitem[Norris \& Norris, 2009]{norrisbook}Norris, R.P., \& Norris, Cilla M., 2009, {\it Emu Dreaming: An Introduction to Australian Aboriginal Astronomy}, Sydney: Emu Dreaming Press. 

\bibitem[Norris \& Hamacher, 2009]{norris09}Norris, R.P. \& Hamacher, D.W., 2009, {\it The Astronomy of Aboriginal Australia}.  In {\it The Role of Astronomy in Society and Culture}, D. Valls-Gabaud \& A. Boksenberg, eds., Cambridge University Press, pp. 39-47 

\bibitem[Norris \& Hamacher, 2011]{norris11}Norris, R.P. \& Hamacher, D.W., 2011, Astronomical Symbolism in Australian Aboriginal Rock Art, {\it Rock Art Research}, Vol.~28(1): p.~99-106, 28, 99  

 
\bibitem[Palmer, 1977]{Palmer} Palmer, K., 1977, Stone Arrangements and Mythology.  {\it Mankind}, Vol. 11, No. 1, pp. 33--38.

\bibitem[Richards, 1996]{richards96}Richards, T. \& Jordan, J., 1996, {\it Aboriginal Archaeology of the Moorabool Basin}, Unpublished Report to Aboriginal Affairs Victoria.

\bibitem[Stanbridge, 1861]{stanbridge}Stanbridge, W.E., 1861, Some particulars of the general characteristics, astronomy, and mythology, of the tribes in the central part of Victoria, Southern Australia.  {\it Transactions of the Ethnological Society of London}, Vol 1, pp. 286--303.

\bibitem[Towle, 1939]{Towle} Towle, C.C., 1939, Stone arrangements and other relics of the Aborigines found on the lower Macquarie River, N.S.W., at and near Mount Foster and Mount Harris.  {\it Mankind}, Vol 2, pp. 200--209.

\end{thebibliography}

\clearpage

\end{document}